\documentclass[aps,superscriptaddress,nofootinbib,eqsecnum,prd,notitlepage,twocolumn,floatfix]{revtex4} 

\pdfoutput=1 

\usepackage{amssymb}   
\usepackage{amsmath}   
\usepackage{graphicx}   
\usepackage{subfigure}
\usepackage{hyperref}
\usepackage{float}
\usepackage[cmyk]{xcolor}

\newcommand{\PR}[1]{\ensuremath{\left[#1\right]}}
\newcommand{\PC}[1]{\ensuremath{\left(#1\right)}}

\everymath{\displaystyle}
\newcommand{\limit}[3]
{\ensuremath{\lim_{#1 \rightarrow #2} #3}}

\begin{document}

\title{Projected Proca Field Theory: a One-Loop Study}

\author{R.F. Ozela} 
\affiliation{Institute for Theoretical Physics, Utrecht University,
  3584 CC Utrecht, The Netherlands}
\affiliation{Faculdade de F\'{\i}sica, Universidade Federal do Par\'a,
  66075-110 Bel\'em, PA,  Brazil}

\author{Van S\'ergio Alves} 
\affiliation{Faculdade de F\'{\i}sica, Universidade Federal do Par\'a,
  66075-110 Bel\'em, PA,  Brazil}

\author{E.C. Marino} 
\affiliation{Instituto de F\'{\i}sica, Universidade Federal do Rio de Janeiro,
21941-972 Rio de Janeiro, RJ , Brazil}

\author{Leandro O. Nascimento} 
\affiliation{Faculdade de F\'{\i}sica, Universidade Federal do Par\'a,
  68800-000 Breves, PA,  Brazil} 
  
\author{J.F. Medeiros Neto} 
\affiliation{Faculdade de F\'{\i}sica, Universidade Federal do Par\'a,
66075-110 Bel\'em, PA, Brazil}

\author{Rudnei O.  Ramos} 
\affiliation{Departamento de F\'{\i}sica Te\'{o}rica, Universidade do
  Estado do Rio de Janeiro, 20550-013 Rio de Janeiro, RJ, Brazil}

\author{C. Morais  Smith} 
\affiliation{Institute for Theoretical Physics, Utrecht University,
  3584 CC Utrecht, The Netherlands}


\begin{abstract}
 
The recent discovery of two-dimensional Dirac materials, such as
graphene and transition-metal-dichalcogenides, has raised questions
about the treatment of hybrid systems, in which electrons moving in a
two-dimensional plane interact via virtual photons from the
three-dimensional space. In this case, a projected non-local theory,
known as Pseudo-QED, or reduced QED, has shown to provide a correct
framework for describing the interactions displayed by these systems.
In a related situation, in planar materials exhibiting
a superconducting phase, the electromagnetic field has a typical exponential
decay that is interpreted as the photons having an effective
mass, as a consequence of the Anderson-Higgs mechanism. Here, we use an analogous projection to
that used to obtain the pseudo-QED to derive a Pseudo-Proca
equivalent model. In terms of this model, we unveil the main effects
of attributing a mass to the photons and to the quasi-relativistic
electrons. The one-loop radiative corrections  to the electron mass,
to the photon and to the electron-photon vertex are computed. 
{We calculate the quantum corrections to the electron $g$-factor and show that it smoothly goes to zero in the limit when the photon mass is much larger than the electron mass. In addition, we correct the results obtained for graphene within Pseudo-QED in the limit when the photon mass vanishes.}

\end{abstract}


\maketitle

\section{Introduction}

The use of  Quantum Field Theories (QFT) to describe two-dimensional
systems has gained increased attention during the last years. This is
due to the great agreement obtained between the theoretical
predictions and the experimental data in many condensed-matter
systems. Examples range from the integer and the fractional quantum
Hall effects~\cite{Ando,Tsui,anyons,Fradkin,Heinonen} to the  study of
transport in
graphene~\cite{Herbut,Voz1,Voz2,Gusynin,HerbJuri,Gorbar,Semenoff,Bfraco,PRX},
excitonic properties of {Trasition Metal Dichalcogenides}
(TMDs)~\cite{TMDs,exc,Nuno} and superconductivity in layered
materials~\cite{Zhang,Tesanovic,Franz,Kilveson,filmes}. 

Among those, a particular interest is devoted to the so-called Dirac
systems, which exhibit quasi-relativistic  dynamics, with massless
(e.g., graphene) or massive (e.g., silicene and TMDs) electrons moving with
the {}Fermi velocity.~These Dirac materials then operate as QFT
laboratories~\cite{castroneto}.

The description of the electron-electron interactions in planar
materials is rather involved because the photons mediating the
interactions live in (3+1)D, whereas the electron-dynamics is
constraint to a two-dimensional spatial plane. The appropriate theory
to capture this electron-photon dimensional mismatch is the so called
pseudo-quantum electrodynamics (PQED)~\cite{Marino1} (sometimes also
named reduced-QED~\cite{Gusynin1,teber,teber1}). In this approach, a
projected electromagnetic field emulates the properties of the (3+1)D photons.

The PQED has been demonstrated to be unitary~\cite{unitariedade}, causal~\cite{amaral} and
has been successfully applied to describe several properties of
very thin systems (graphene-like structures)~\cite{PRX, TMDlfran}, where the 2d approximation is a reasonable assumption. Among others, we highlight the Fermi velocity renormalization in the absence~\cite{Voz2} or in the
presence~\cite{Bfraco}~of a magnetic field in the vicinity of a
conducting plate~\cite{placa} or in a cavity~\cite{cavity}. In
addition, it provided a theoretical description of the  Quantum Valley
Hall Effect, quantum corrections for the longitudinal conductivity in
graphene~\cite{PRX}, and of the corrections to the  electron's
\textit{g}-factor due to interactions~\cite{fator-g}. When  accounting
for massive electrons, the theory was shown to describe the excitonic
spectrum in  TMDs~\cite{exc}~and dynamical chiral symmetry
breaking~\cite{Juricic}.  A dual  PQED type of model describing the
interaction between point vortex excitations and with some interesting
properties has also been recently constructed~\cite{Alves:2019bqt}.

Another topic that attracted much attention recently is the
superconductivity that shows up at low temperatures in bilayer
graphene twisted at the \textit{magic} angle
$\left(\theta\approx1.1^{\circ}\right)$~\cite{bilayer,banda}.
An interesting related question is how to treat a decaying magnetic field
due to the presence of superconductivity in that case. The
characteristic Meissner-Ochsenfeld magnetic field screening in
superconductors is usually described by  massive photons (e.g., as
described by using the Ginzburg-Landau equations in
three-spatial dimensions). Such a model, however, lacks the projection
component to describe planar relativistic condensed-matter systems,
which are accounted for within the PQED.

Here, we will follow steps similar to the ones that led to the
construction of the PQED model~\cite{Marino1} and develop a theory to
describe electron-electron interactions through a \textit{massive} vector (Proca) field.
We consider the Proca model without concern as to the origin of the mass term,
which in a more
fundamental theory comes from a Higgs-like mechanism.
 For a derivation of the model we use here, see Ref.~\cite{Gabriel}, which also shows that a Yukawa potential is generated in the static limit. 
Such model constrains only the matter (electrons)
current to the spatial $xy$ plane. The corresponding quantum partition functional
is defined initially in 3+1-dimensions and then the third spatial
dimension is integrated out. This procedure is very much
analogous to the one that links the (3+1)D Maxwell model to
the 2+1-dimensions PQED model. 

This work is organized as follows. In Sec.~\ref{MP-PlanarCurent} we
present the model used in this work and we derive its planar
dimensional reduction  in a procedure analogous to that used to derive
the PQED model.  In Sec.~\ref{FRules}, we compute the electron and
photon self-energies for the model, as well as the interaction vertex,
within the leading order (one-loop) level.  In this same section, we
also explicitly derive the \textit{g}-factor for our model. A
comparison of our results with previous ones, when considering the
massless regime, is also performed. In Sec.~\ref{Resultados} we present
our conclusions. Some technical details of the calculation of the
\textit{g}-factor are give in the App.~\ref{app}.

\section{Pseudo Proca model}
\label{MP-PlanarCurent}

Let us first consider the (3+1)D Proca (P) model,
including the coupling  to a general conserved current $J^{\mu}$. The quantum partition functional is
\begin{equation}
{ Z}_{\mathrm{P}}[J^{\mu}]  =  \int {\cal D}A_{\mu} \exp\left( i
S_{\mathrm{P}} \right), 
\label{Zfunc}
\end{equation}
where $A_\mu$ is a (3+1)D massive vector field and $S_{\mathrm{P}}$ is
the action, given by
\begin{equation} 
\!\!\!\!S_{\mathrm{P}} \!=\! \int d^4 \chi  \left(
-\frac{1}{4}F_{\mu\nu} F^{\mu\nu} + \frac{m^2}{2}A^\mu A_\mu - eA_\mu
{J}^{\mu}  \right),
\label{action}
\end{equation}
where $F_{\mu\nu} = \partial_\mu A_\nu - \partial_\nu A_\mu$ and $m$ is the vector field mass. Natural units ($\hbar=c=1$) are considered for the
remainder of this section. The vector field propagator is directly
derived from Eq.~(\ref{action}), 
\begin{equation}
\!\!\!\!G^{\mu\nu}_{\mathrm{P}}(\chi - \chi')=  \int
\frac{d^4k}{(2\pi)^4}\frac{e^{-ik\cdot(\chi - \chi')}}{k^2 - m^2 }
\left[\eta^{\mu\nu}-\frac{k^\mu k^\nu}{m^2}\right] ,
\label{propagatorMP}
\end{equation}
with $\chi$ and $\chi'$ representing points in the (3+1)D
spacetime and $\eta^{\mu\nu}=\text{diag}(+,
-, -, -)$ is the metric tensor.  Integrating out the vector field in Eq.~(\ref{Zfunc}), we obtain
\begin{eqnarray}
\hspace{-7mm} { Z}_{\mathrm{P}}[J^\mu]  &=&  {\cal Z}_0
e^{-i\tfrac{e^2}{2} \int d^4\chi\, d^4\chi'\, J_{\mu}(\chi)
  G^{\mu\nu}_{\mathrm{P}}(\chi - \chi') J_{\nu}(\chi')} ,
\label{ZMP}
\end{eqnarray}
where ${\cal Z}_0$ is a normalization constant, independent of
$J^\mu$. Note that by the conservation of the current, $\partial_\mu
J^\mu=0$, the last term in Eq.~(\ref{propagatorMP}) ($k^\mu
k^\nu/m^2$) vanishes in Eq.~(\ref{ZMP}). 

By using the constrain of having only currents in the $xy$ plane, we
can explicitly write the current $J^\mu$ as
\begin{equation}
J^{\mu}(\chi) = \left\{
\begin{array}{cc}
{\cal J}^{\hat{\mu}}(t,\, x,\, y)\delta(z), & \mbox{if } \mu = 0,\,
1,\, 2\, ,\\ 0, & \mbox{if } \mu = 3,\, \ \, \; \ \, \  \ \  \label{j4-3}
\end{array}
\right.
\end{equation}
where the hat over an index notation is used to identify objects that
assume three values, i.e., $\hat \mu=0,1,2$. 

After the integration in $z$ and $z'$ space coordinates,
Eq.~(\ref{ZMP}) can be written as
\begin{eqnarray}
\hspace{-6mm} {\cal Z}_{\mathrm{PP}}[{\cal J}]  &=&  {\cal Z}_0 \, e^{-i
  \tfrac{e^2}{2} \int d^3\zeta\, d^3\zeta' J_{\hat{\mu}}(\zeta) {\cal
    G}^{\hat{\mu}\hat{\nu}}_{\mathrm{P}}(\zeta - \zeta')
  J_{\hat{\nu}}(\zeta') } ,
\end{eqnarray}
with  $\zeta$ and $\zeta'$ denoting points in the 2+1-dimensions
spacetime and ${\cal G}^{\hat{\mu}\hat{\nu}}_{\mathrm{P}}(\zeta -
\zeta')$ is given by
\begin{equation}
{\cal G}^{\hat{\mu}\hat{\nu}}_{\mathrm{P}}(\zeta - \zeta') = i\int
\frac{d^4k}{(2\pi)^4} \frac{\eta^{\hat{\mu}\hat{\nu}}}{k^2 - m^2}
\left.e^{-ik\cdot(\chi - \chi')}\right\vert_{z = z' = 0},
\label{propMPPlane}
\end{equation}
where the momentum integration is over the energy-momentum four-vector
$k^\mu$.

If we now perform the integration over the third component of $k^\mu$, i.e.,
$k_z$, in Eq.~(\ref{propMPPlane}) and hence restrict the dynamics to the $xy$-space plane, we obtain
\begin{equation}
{\cal G}^{\hat{\mu}\hat{\nu}}_{\mathrm{P}}(\zeta - \zeta') 
= \frac{i}{2}\int \frac{d^3 k} {(2\pi)^3}\frac{ \eta^{\hat{\mu}\hat{\nu}} 
e^{-i k\cdot (\zeta-\zeta')}} {\sqrt{k^2 - m^2}},\label{propMPPlane2}
\end{equation}
with $k$ now indicating a three-vector, stressing
the fact that we are now working in the reduced space (effectively, in
2+1-dimensions).
However, the above result can also be obtained if we start
from a completely 2+1-dimensions, in principle nonlocal, model from the very beginning. We will name this model the ``Pseudo Proca'' (PP) model, in analogy with the  case of the PQED model. Notice that the denominator of the propagator in Eq.~(\ref{propMPPlane2}) is  significantly different from the original Proca model and this impacts directly the quantum corrections in a mixed dimensionality  physical system.

The Lagrangian density for this model is then expressed as
\begin{eqnarray}
{\cal L}_{\mathrm{PP}} & = & -\frac{1}{2}
\frac{F_{\hat{\mu}\hat{\nu}}F^{\hat{\mu}\hat{\nu}}} {\sqrt{-\Box -
    m^2}}  - eA_{\hat{\mu}} {\cal J}^{\hat{\mu}} -\frac{m^2}{2}
\frac{A^{\hat{\mu}}A_{\hat{\mu}}}{\sqrt{-\Box - m^2}}, \nonumber\\
\label{SPMP}
\end{eqnarray}
with ${\cal J}^{\hat{\mu}}$ being the general 2+1-dimensions current
defined in Eq.~(\ref{j4-3}) and $\Box$  the d'Alembertian operator
(which must be understood as a convolution).  The free propagator
associated with ${\cal L}_{\mathrm{PP}}$ is simply given by 
Eq.~(\ref{propMPPlane2}), as can be easily verified.
Thus, we immediately realize that 
\begin{equation}
{\cal G}^{\hat{\mu}\hat{\nu}}_{\mathrm{PP}}(\zeta-\zeta') =
G^{\hat{\mu}\hat{\nu}}_{\mathrm{P}}(\chi-\chi')\Bigr|_{z = z' = 0}\,
\end{equation}
and, therefore, the quantum partition function of the Pseudo-Proca and that of the (3+1)D Proca models are
completely equivalent, as long as the currents of the latter are
constrained to a plane, such as in Eq.~(\ref{j4-3}).

\section{Radiative corrections}
\label{FRules}

In this section, we consider a soft symmetry-breaking term in the
Dirac action through the {}Fermi velocity $v_F$, in order to reproduce
the Dirac-like low-energy electronic dispersion. The Lagrangian
density in the now 2+1-dimensional Minkowski space is then
given by
\begin{eqnarray}
    {\cal L}&=& -\frac{1}{2}  \frac{F_{\mu\nu}F^{\mu\nu}} {\sqrt{-\Box
        - m^2}} -\frac{m^2}{2}  \frac{A^{\mu}A_{\mu}}{\sqrt{-\Box -
        m^2}} \nonumber \\ &+&\bar{\psi}(i \gamma^{0}\partial_{0} +i
    v_{F}\gamma^{i} \partial_{i} - Mc^2)\psi
    -e\,\bar{\psi}\gamma^{0}\psi A_{0} \nonumber \\ &-&
    e\,\beta\bar{\psi}\gamma^{i}\psi A_{i}, 
\end{eqnarray}
with $\psi$ a four-component Dirac spinor, $\beta =v_F/c$ and
$M$ the Dirac fermion mass. In the above equation and  from this
point on, we have explicitly retrieved the dimension of the speed of
light $c$ (but still keeping $\hbar =1$ units). {}For convenience, and
to avoid overloading the notation, we suppress the hat of the Lorentz
index. 

The {}Feynman rules for the model are obtained as
usual~\cite{Peskin}.~The interaction vertex is given by
$\Gamma^{\alpha}=-ie\left(\gamma^0,\beta \, \gamma^i \right)$, the
fermion propagator is
\begin{equation}
 S_F(p) =i \left( \dfrac{\gamma^0 p_0 + v_F \gamma^i p_i + M
   c^2}{p_0^2-v_F^2 {\bf p}^2-M^2 c^4} \right),
\label{propfermion}
 \end{equation}
and the massive vector field propagator reads
\begin{equation}
\Delta_{\mu\nu} (p) =-\frac{i c \eta_{\mu\nu}} {2\sqrt{p_0^2 - c^2 \,
    {\mathbf p}^2 - m^2 c^4}}.
\end{equation}

\subsection{Electron self-energy}
\label{photon_mass_in_electron_self_energy}

\begin{figure}[!htb]
    \centering \includegraphics[width=6cm]{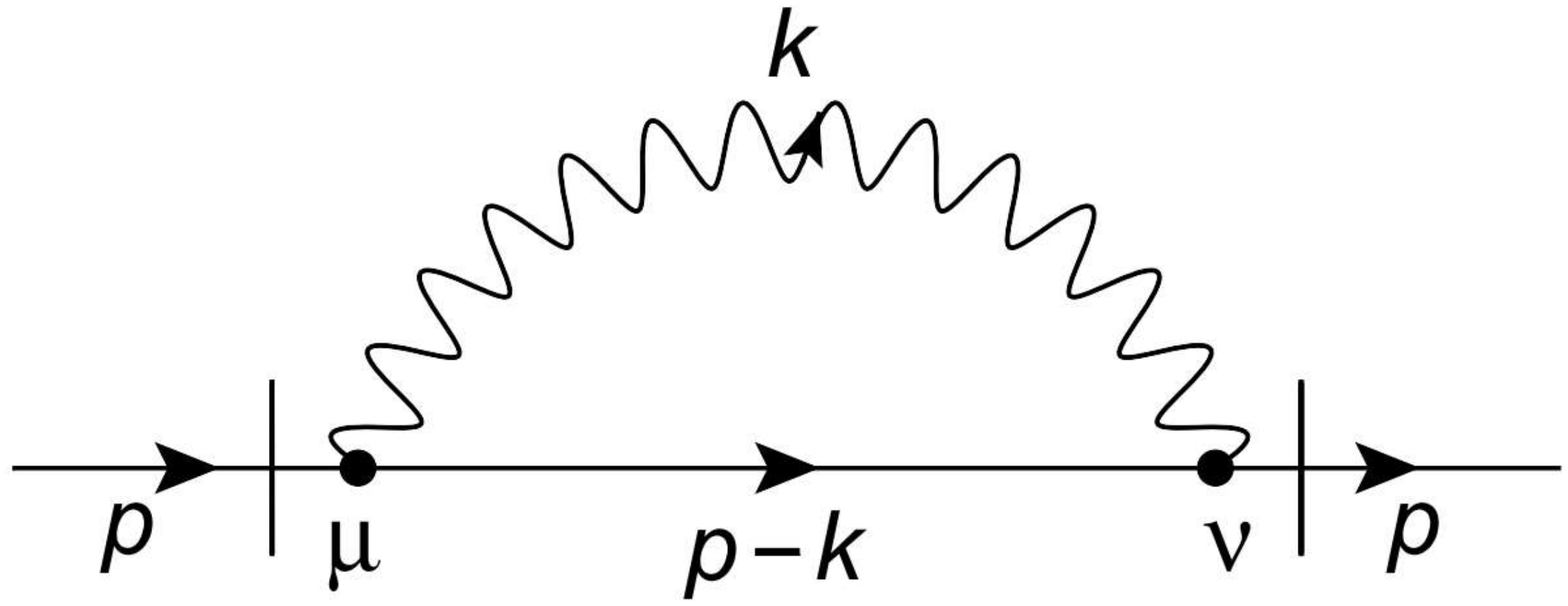}
    \caption{Feynman Diagram for the one-loop correction to the fermion
      self-energy.}
    \label{self-e}
\end{figure}

The electron self-energy at one-loop order is shown in
{}Fig.~\ref{self-e} and the regularized amplitude is given by 
\begin{eqnarray}
-i\Sigma(p) = \int \frac{d^D k}{(2\pi)^D} \Gamma^{\mu}  S_F(p-k)
\Gamma^{\nu} \Delta_{\mu\nu}(k),\label{Sigma1_a}
\end{eqnarray}
where $D=d+1$.  In what follows, all momentum integrations in the loop
radiative quantities are computed using standard dimensional
regularization procedure~\cite{Bollini,thooft}, with $e \to
\mu^{\epsilon/2} e$, $d=2-\epsilon$ and  $\mu$ is the dimensional
regularization energy scale. 

By substituting the propagators and vertices in Eq.~(\ref{Sigma1_a}),
we obtain %
\begin{equation}
-i\Sigma (p)  = \frac{(ie)^2 \mu^\epsilon c}{2}  \int  \frac{d^D
  k}{(2\pi)^D}  \dfrac{N_1}{D_1} ,
\label{integralinicial}
\end{equation}
where $N_1$ and $D_1$ are given, respectively, by
\begin{eqnarray} 
 N_1  &=& v_F(p-k)_i  \left(\gamma^0 \gamma^i \gamma_0
 +\beta^2\gamma^j  \gamma^i \gamma_j \right) \nonumber\\ &+& (p-k)_0
 \left(\gamma^0 \gamma^0 \gamma_0 +\beta^2\gamma^j \gamma^0 \gamma_j
 \right)  \nonumber \\ &+& M c^2 \left(\gamma^0  \gamma_0
 +\beta^2\gamma^j \gamma_j \right)  , \\ \nonumber \\ D_1 &=&
 \left[(p-k)_0^2-v_F^2 ({\bf p}-{\bf k})^2-M^2 c^4\right] \nonumber
 \\ &\times& \sqrt{k_0^2 - c^2 \, {\bf k}^2 - m^2 c^4} .
  \end{eqnarray}

Using, hereinafter, the gamma matrix properties
$\gamma^{0}\gamma_{0}=1$ and $\gamma^{j}\gamma_{j}=2$, we have that
\begin{widetext}
\begin{equation} \label{sigmaantesdeintegrar}
-i\Sigma (p)=  \frac{(ie)^2\mu^\epsilon c} {2} \int \frac{dk_0}{2 \pi}
\frac{d^d k}{(2\pi)^d}  \dfrac{\gamma^0 (p-k)_0\left(
  1+2\beta^2\right)  - v_F \gamma^i (p-k)_i  + \left( 1 + 2\beta^2
  \right) M c^2} {[(p-k)_0^2-v_F^2 ({\mathbf p}-{\mathbf k})^2-M^2
    c^4] (\sqrt{k_0^2 - c^2 \, {\mathbf k}^2 - m^2 c^4})}.
\end{equation}
\end{widetext}
Using the {}Feynman parameterization,
\begin{equation}
    \frac{1}{a\sqrt{b}}=\dfrac{3}{4} \int_0^1 dx\left\{
    \frac{(1-x)^{-1/2}}{{[a\,x + (1-x)\, b]^{3/2}}} \right\},
\end{equation}
and performing the integration in $k_0$, we obtain that 
Eq.~(\ref{sigmaantesdeintegrar}) can be rewritten as
\begin{equation}
-i\Sigma (p)  =\frac{(ie)^2\mu^\epsilon c}{4\pi} \int_0^1
\dfrac{dx}{\sqrt{1-x}} \int \frac{d^d  k}{(2\pi)^d} \dfrac{N_2}{D_2},
\end{equation}
where $N_2$ and $D_2$ are defined, respectively, as 
\begin{eqnarray} 
N_2 &=& \left( 1 + 2\beta^2 \right) M c^2 +\left(
1-2\beta^2\right)\gamma^0 p_0 (1-x) \nonumber\\ &-& \left(1-{\frac
  {v^2_F x}{\cal B}} \right) v_F \gamma^i p_i,
\label{N2}
\\ {D_2} &=& {\cal B} \left[ \left( {\bf k}-\frac {{\bf p} v^2_F
    x}{{\cal B}} \right) ^{2}- {\frac {{\bf p}^{2}{{
          v_F}}^{4}{x}^{2}}{{\cal B}^{2}}}+ {\frac {\cal A}{\cal B}}
  \right],
\end{eqnarray}
with ${\cal A}$ and ${\cal B}$ defined as
\begin{eqnarray} 
{\cal A}&=& p^2_0 {x}^{2}+ \left[ ({M}^{2}-m^2){c}^{4}+{\bf p}^{2}{{
      v_F}}^{2}-p^2_0 \right] x  + {m}^{2}{c}^{4},\nonumber \\
\\  {\cal B}&=& c^2 \left[1- x\, (1-\beta^2)\right].
\end{eqnarray}
The electron self-energy then becomes,
\begin{equation}
-i\Sigma (p) = \frac{(ie)^2\, c\mu^{\epsilon}}{4\pi} \int_0^1
\dfrac{dx}{{\cal B}\sqrt{1-x}}  \int \frac{d^d k}{(2\pi)^d}
\dfrac{N_2} {  k^2-\Delta} ,
\label{shiftada}
\end{equation}
where 
\begin{equation}
\Delta =  {\frac {{\bf p}^2{{ v_F}}^{4}{x}^{2}}{{\cal B}^{2}}}+{\frac
  {\cal A}{\cal B}}.
\end{equation}

Performing the integration in the arbitrary dimension $d=2-\epsilon$
in Eq.~(\ref{shiftada}) in the dimensional regularization procedure,
we obtain
\begin{eqnarray}
\Sigma (p)&=&\frac{(ie)^2\, c\mu^{\epsilon}}{16 \pi^2} \int_0^1
\dfrac{dx}{\sqrt{1-x}}  \frac {\Gamma \left(\tfrac{\epsilon}{2}
  \right)\,N_2  } { {\cal B} } \left[\dfrac{ \Delta}{4\,\pi
  }\right]^{-\tfrac{\epsilon}{2}} \nonumber \\ &=& \frac{(ie)^2\,
  c}{16 \pi^2} \int_0^1  \dfrac{dx}{\sqrt{1-x}}  \frac {N_2  } { {\cal
    B} } \nonumber \\ &\times& \left[ \frac{2}{\epsilon}-\gamma_E -\ln
  \left(\frac{\Delta}{4 \pi \mu^2}\right)  +{\cal O}(\epsilon) \right]
\nonumber \\ &=& \Sigma_{\rm finite} (p) + \Sigma_{\rm div}(p),
\end{eqnarray}
where
\begin{eqnarray}
\Sigma_{\rm finite} (p)&=& -\frac{(ie)^2\, c}{16 \pi^2} \int_0^1
\dfrac{dx}{\sqrt{1-x}}  \frac {N_2  } { {\cal B} }  \nonumber
\\ &\times& \left[\gamma_E +\ln \left(\frac{\Delta}{4 \pi
    \mu^2}\right) \right]
\label{Sigmafin}
\end{eqnarray}
and
\begin{eqnarray}
\Sigma_{\rm div} (p)= \frac{(ie)^2\, c}{8 \pi^2 \epsilon} \int_0^1
\dfrac{dx}{\sqrt{1-x}}  \frac {N_2  } { {\cal B} }
\label{Sigmadiv}
\end{eqnarray}
are the finite and divergent contributions to the electron
self-energy, respectively, in the pseudo-Proca QED.  One notices that
the divergent part of the electron self-energy, Eq.~(\ref{Sigmadiv}),
is independent of the vector field mass and it is in fact exactly the same
result as that obtained in the PQED
case~\cite{livroMarino}. Consequently, the Fermi velocity
renormalization will also be the same.

{}Substituting Eq.~(\ref{N2}) in Eq.~(\ref{Sigmafin}), we immediately
read the one-loop contributions to the electron mass and the
wave-function renormalization terms. In particular, the one-loop
correction for the fermion mass at zero-external momenta is
\begin{eqnarray}
&\hspace{-11mm}\Sigma_{\rm finite}^M (0)= &\hspace{-9mm} \frac{e^2\, c}{16 \pi^2} \left( 1 +
2\beta^2 \right) M  \int_0^1  \dfrac{dx}{\sqrt{1-x}}  \,   \nonumber \\
&\frac{1  }
{ 1- x\, (1-\beta^2) }& \left\{\gamma_E +\ln
\left[\frac{M^2c^2 x+ m^2 c^2(1-x)}{4 \pi \mu^2 [1-x\,
      (1-\beta^2)]}\right] \right\}.\nonumber\\
\label{SigmaM}
\end{eqnarray}
The PQED result is obtained by simply setting $m=0$ in
Eq.~(\ref{SigmaM}).

\begin{center}
\begin{figure}[!htb]
\includegraphics[width=8.2cm]{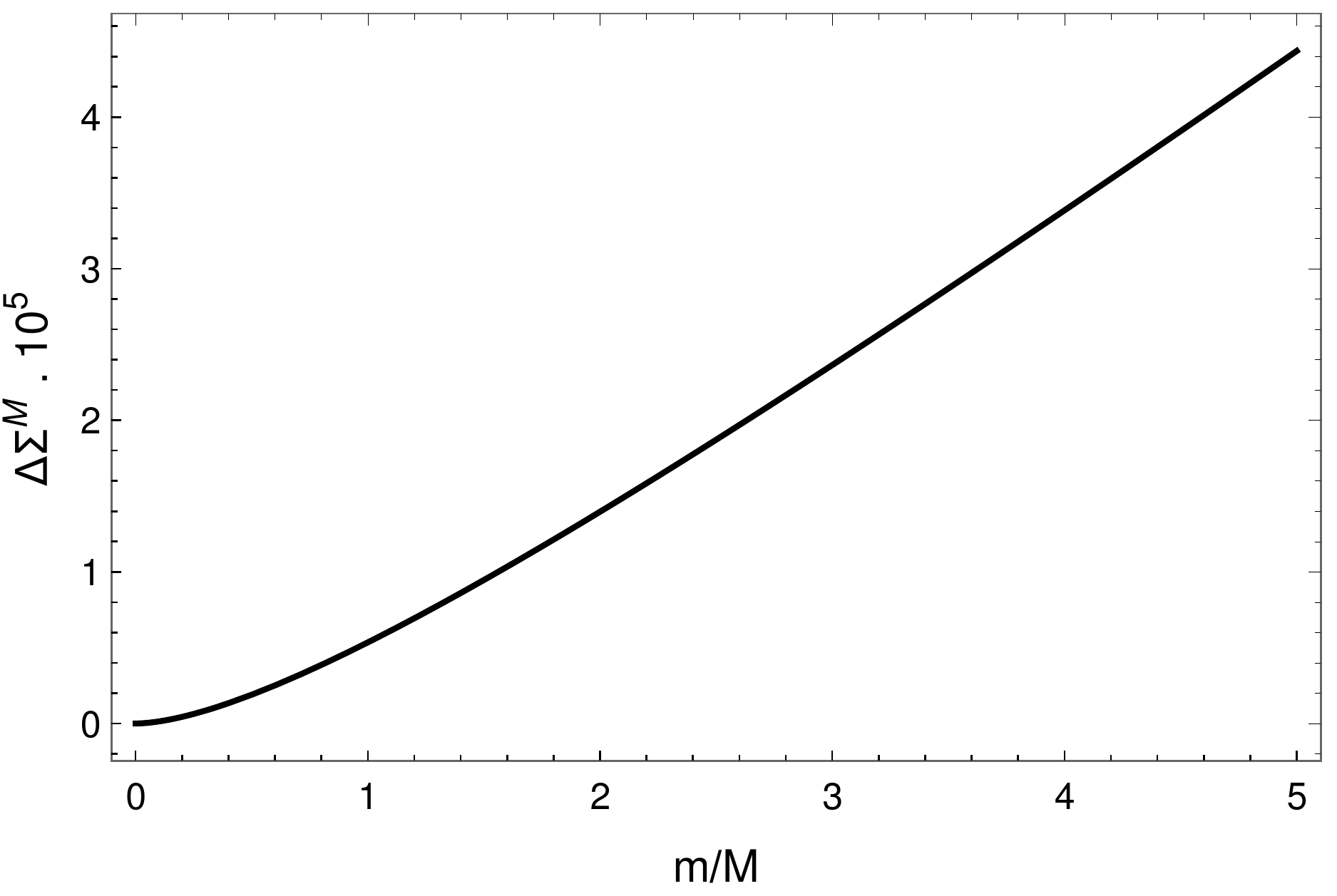}
\caption{The difference of the self-energy correction for the fermion
  mass, with respect to the PQED one, as a function of the mass ratio
  ($m/M$).  {}For illustrative purposes only, the {}Fermi velocity was taken to be $v_F=c/300$ 
  and we also have used $e^2= 4 \pi v_F \alpha$, where $\alpha=1/137$ is the fine-structure
  constant.}
\label{Meff}
\end{figure}
\end{center}

In {}Fig.~\ref{Meff} we show the behavior of the ratio $\Delta \Sigma^M\equiv [ \Sigma_{\rm finite}^M (0) - \lim_{m\to 0}\Sigma_{\rm finite}^M (0)]/(M c^2)$ as a function of ($m/M$). We note that the electron self-energy correction increases with the vector field mass.

\subsection{The vector field self-energy}

\begin{figure}[!htb]
    \centering \includegraphics[width=5cm]{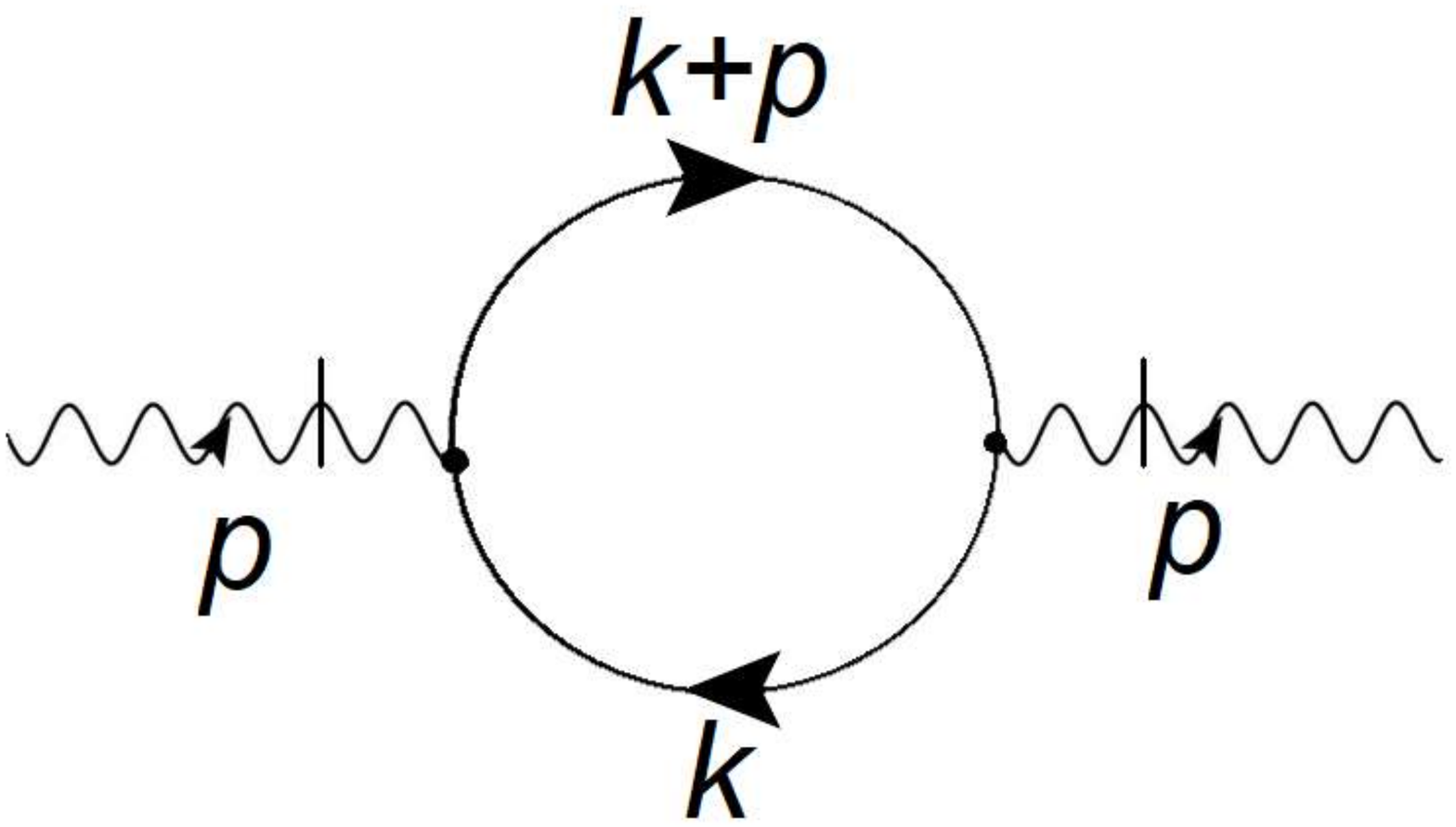}
    \caption{The one-loop vector field vacuum polarization diagram.}
    \label{polariz}
\end{figure}

Let us now compute the vector field self-energy, shown in  {}Fig.~\ref{polariz}, 
for the present model.  Explicitly, we have that
\begin{eqnarray}
&& i\Pi^{\mu\nu}(p_0,\textbf{p}) = -\int\frac{d^D k}{(2\pi)^{D}}
  \texttt{Tr} \Big[ \Gamma^{\mu} S_F(k+p)\Gamma^{\nu} S_F(k)\Big].
  \nonumber \\
\label{photonS}
\end{eqnarray}

The manipulation of the Dirac algebra proceeds in a standard
fashion~\cite{Peskin}. Separating the components of the polarization
tensor and performing the integration over $k_{0}$ and $\textbf{k}$ in
Eq.~(\ref{photonS}), we can write
\begin{eqnarray}
\!\!\!\!\!\!    \Pi^{00}(\bar p)\!\! &=& 
    \!\!\frac{e^{2}}{8\pi v_{F}} \left[I_{1}+
      \frac{(\bar{p}^{2}-2p_{0}^2)I_{2}}{v_{F}}\right.\left. + 
      \frac{Mc^2 I_{3}}{v_{F}^{2}} \right], 
\\ \!\!\!\!\!\! \Pi^{ij}(\bar p) &=&
    \frac{e^{2}v_F}{8\pi c^{2}}  \left[
      \PC{\frac{\eta^{ij}\bar{p}^2}{v_{F}^2}-2p^{i}p^{j}}I_{2}
      \right. \nonumber \\          &+&
      \left.\eta^{ij}\PC{I_{1}+\frac{Mc^2 I_{3}}{v_{F}^2}}\right] ,
\\ \!\!\!\!\!\! \Pi^{0j}(\bar p)&=& -\frac{e^{2}}{4\pi v_{F} c} p^{0}p^{j}
    I_{2},
\end{eqnarray}
where for convenience we have introduced the notation~$\bar
p=\sqrt{p^2_0-v^2_F{\bf p}^2}$ and $I_1$, $I_2$ and $I_3$ are given,
respectively, by
\begin{eqnarray}
I_{1}&=&\frac{1}{v_{F}}\int_{0}^{1} dx \sqrt{x(1-x)\bar p^2- M^2 c^4}
\nonumber \\ &=&\frac{-i}{8v_{F}\bar p} \left\{ 4 Mc^2 \bar p + \PR{4
  M^2c^4-\bar p^2}\ln{\left(\frac{2 Mc^2+\bar p}{{2Mc^2-\bar p}}\right)}
\right\} ,
\nonumber\\ 
\\
I_{2}&=&v_{F}\int_{0}^{1}dx\frac{x(1-x)}{\sqrt{x(1-x)\bar p^2 -
    M^2c^4}} \nonumber\\ &=&\frac{i\,v_{F}}{8\bar p^{3}} \left\{
4 Mc^2\bar p + \PR{4 M^2c^4+\bar p^2}\ln{\PC{\frac{2Mc^2+\bar
      p}{2Mc^2-\bar p}}} \right\}  , 
\nonumber \\ \\ 
I_{3} &=&
\int_{0}^{1} \frac{ v_{F}\ dx}{ \sqrt{ x(1-x)\bar p^2 - M^2c^4}}  =
\frac{i\,v_{F}}{\bar p}\ln{\left(\frac{2Mc^2+\bar p}{2Mc^2-\bar
    p}\right)} .  \nonumber \\
\end{eqnarray}

Notice that the use of the dimensional regularization scheme makes the
vector field self-energy explicitly finite.  {}Furthermore, as
at the one-loop level the vector field self-energy involves only the
electron propagator, the result turns out to be identical to that
computed in the context of the 2+1-D $QED$ at one-loop order.

\subsection{Vertex correction and the \textit{g}-factor}
\label{photon_mass_in_vertex_correction}

\begin{figure}[!htb]
    \centering \includegraphics[height=50mm]{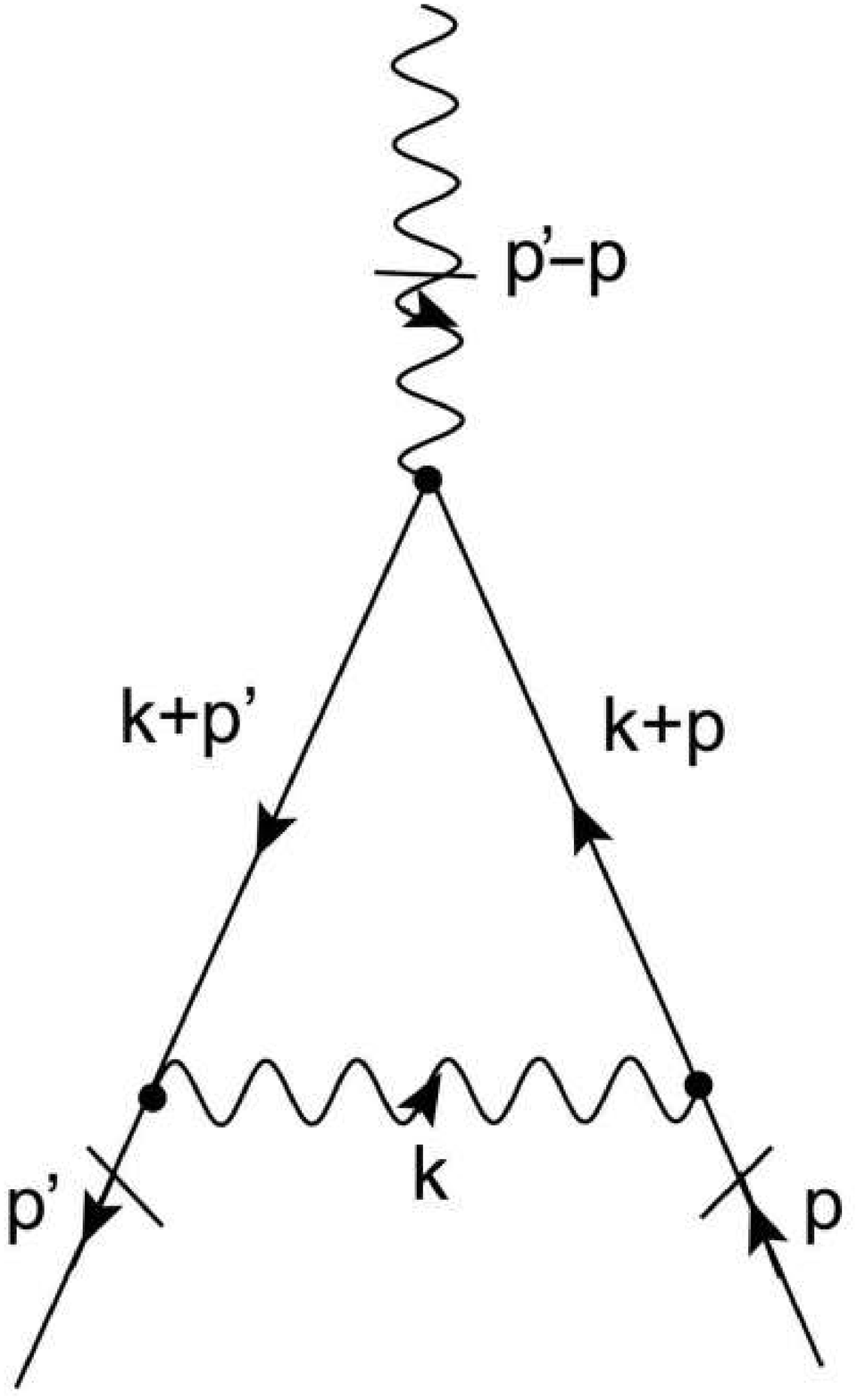}
    \caption{Feynman diagram for the one-loop correction to the
      interaction vertex.}
    \label{fluxos}
\end{figure}

To have a complete one-loop analysis of relevant quantities, we also compute next
the one-loop interaction vertex radiative contribution
$\Gamma^{\nu}_{\rm 1L}$ as shown in {}Fig.~\ref{fluxos}.

Explicitly, $\Gamma^{\nu}_{\text 1L}$ is given by
\begin{equation}\label{vertexStruc}
\Gamma^{\nu}_{\rm 1L} =  \int \frac{d^D k}{(2\pi)^D}
\Gamma^{\alpha}S_F(k + p') \Gamma^{\nu} S_F(k + p)
\Gamma^{\beta}\Delta_{\alpha\beta}(k).
\end{equation}

We will analyze the quantity $M^{i} = \overline{u}(p')\Gamma^{i}
u(p)$, which relates to the two external fermion lines, $\Bar{u}(p')$
and $u(p)$, in the spatial component of the vertex diagram (because
the temporal part of it does not affect the \textit{g}-factor), such
that we can use the Dirac equations to simplify the calculations
further.

When represented according to the flux choices in {}Fig.~\ref{fluxos},
and after applying a parameterization detailed in the
Appendix~\ref{app}, the diagram is written  in its parameterized form
as
\begin{equation}\label{vertice}
M^{i} = \Bar{u}(p') \left( \int_0^1{dx} \int_0^{1-x} \dfrac{dy}{6
  \pi^2}  \dfrac{M\, p^i
  v_F^2}{\sqrt{1-x-y}}    \, \dfrac{N_3}{\cal K} \right)\, u(p),
\end{equation}
with
\begin{equation}
N_3 =\frac{ 2(1-x-y) \gamma^j \omega_j p_i v_F^2 }{{\cal D}}
-\frac{[\omega_i-2(x+y)^2 p_i] \beta^4}{{\cal D}^2},
\end{equation}
where ${\cal D}$ in the above equation is defined as
\begin{equation}
{\cal D} =[(1-\beta^2)\,(x+y) -1],
\end{equation}
and
\begin{eqnarray}
{\cal K} &=& {\cal D}^2 \left(M^2 c^4-m^2c^4-\omega_0^2\right) c^2 +
\left[v_F^2 (p'-p)_i^2+p_0^2\right] x y		\nonumber \\	 &+&
\left[(x+y)(1-x-y)c^2\right] (p'+p)_\mu^2 +p_0^2 y^2,
\end{eqnarray}
where $\omega_0 = x\,p_0' + y\,p_0$  and $\omega_i = x\,p' + y\,p$.
Then, we detach the terms proportional to $\sigma^{i \alpha}$ from
Eq.~(\ref{vertice}) using Gordon decomposition to select the terms
$M_g^i$ that are relevant to the \textit{g}-factor  (for more details,
see App.~\ref{app}) 
\begin{eqnarray}
M^{i}_g &=& \Bar{u}(p')\left\{ \int_0^1 dx \int_0^{1-x} \frac{dy}{6
  \pi^2} 	\frac{\;M\, v_F \, \sigma^{i\alpha} (p' -
  p)_\alpha\beta^2\textbf{}}{{\cal K} \sqrt{1-x-y}}\right.
\nonumber \\ &\times& \left.  \left[ {\dfrac{2 (x+y)}{\cal
      D}}-{\dfrac{\beta^2(x+y)^2}{{\cal D}^2}} \right]\right\} u(p),
\label{integF2}
\end{eqnarray}
where the subscribed $g$ index was used to identify that we are only
accounting for the $\sigma^{i\alpha}=(i/2)\left[\gamma_\alpha,\gamma_i\right]$ terms which contribute to the
gyromagnetic factor.

To extract the \textit{g}-factor, however, a few conditions are also
necessary, such as the low-energy approximation ($q^2\rightarrow{0}$)
and the mass-shell condition (which for small energy values reads as
$v_F \gamma^j p_j \cong M c^2$) \cite{Peskin,fator-g}. After taking the above conditions, we
conveniently define a variable 
\begin{eqnarray}
{\cal K'}	&=& \{[(x+y)^2-x-y] v_F^2-[(1-x-y)^2]c^2 \} m^2
c^4\nonumber\\ &+& 2\{[(x+y)^2- x- y] c^2- 2(x+y)^2 v_F^2 \} M^2 c^4,
\nonumber \\
	\end{eqnarray}
so that the relevant parts of the vertex correction $M_g^i$ are
reduced to
 \begin{eqnarray}
M^{i}_g &=& \Bar{u}(p')\left\{ \dfrac{1}{6 \pi^2} \int_0^1 dx
\int_0^{1-x} dy  \dfrac{\,M^2 c^2\, \sigma^{i\alpha} q_\alpha v_F^3 }
    {{\cal K'} \sqrt{1-x-y}} \right.  \nonumber \\ &\times& \left.
    \left[ {2\left(x+y\right) {\cal D}-\left(x+y\right)^2 v_F^2
      }\right]\right\}u(p). 
\end{eqnarray}

We can also identify the  \textit{g}-factor correction form factor
$F_2$ by comparing Eq.~(\ref{integF2}) to 
\begin{equation}
    M^i_g= i\, e\, \beta \, \Bar{u}(p') \left(\dfrac{i}{2Mc^2} \,F_2\,
    v_F \,\sigma^{i\nu} q_{\nu}\right) u(p) .
\end{equation}
{
Therefore, $F_2$ is turned into an integral  related to the one obtained in Ref. \cite{fator-g},
but in the present work a contribution depending on the effective vector field mass $m$ also appears.  We then find
\begin{equation}\label{F2}
    F_2 = -i
\dfrac{\alpha^*\,\beta^3}{2\pi } \,{R},
\end{equation}
with $\alpha^* =e^2/4\pi v_F$ the effective fine-structure constant and
\begin{gather} \displaystyle
{R} ={R}\left(\beta, \frac{m}{M}\right)=
\int\displaylimits_{0}^{1} d x \int\displaylimits_{0}^{1-x}\frac{dy}{\sqrt{1-x-y}
}\nonumber
\\
\frac
{
2 (1+2\beta^2)(x+y)+\beta^2(x+y)^2
}
{\frac{m^2}{M^2}(1-x-y){\cal D}^2
   -     (x+ y)^2 {\cal D}}\label{Rbar}
\end{gather}
 a mass-dependent parametric integral.
}

A few predictions emerge from the result in Eq.~(\ref{F2}) when looking at some of the limits achievable through the tweak of 
mass parameters: 
The mostly intuitive case occurs when the
vector field is heavy with respect to the electron mass,
$m \gg M$, which implies {that the parametric integral assumes the form
\begin{gather} \displaystyle
{R}\Bigr|_{m\gg M}
\simeq
\frac{M^2}{m^2}
\int\displaylimits_{0}^{1} d x \int\displaylimits_{0}^{1-x}\frac{dy(x+y)}{(1-x-y)^{3/2}
}\nonumber\\
\frac
{
 2+4\beta^2+\beta^2(x+y)
}
{[(1-\beta^2)\,(x+y) -1]^2 }
\end{gather}
and, consequently, ${ F_2}=0$, \textit{i.e.}, the \textit{g}-factor vanishes.
}

The second relevant limit happens when $m$ goes
to zero 
in the original unprojected model, which after the projection reproduces the usual PQED behavior, \textit{i.e.},
\begin{equation}
\limit{m}{0} F_2= F_2^\mathrm{PQED} .
\end{equation} 
 \begin{figure}[h!]
    \centering 
   \includegraphics[width=8.6
   cm]{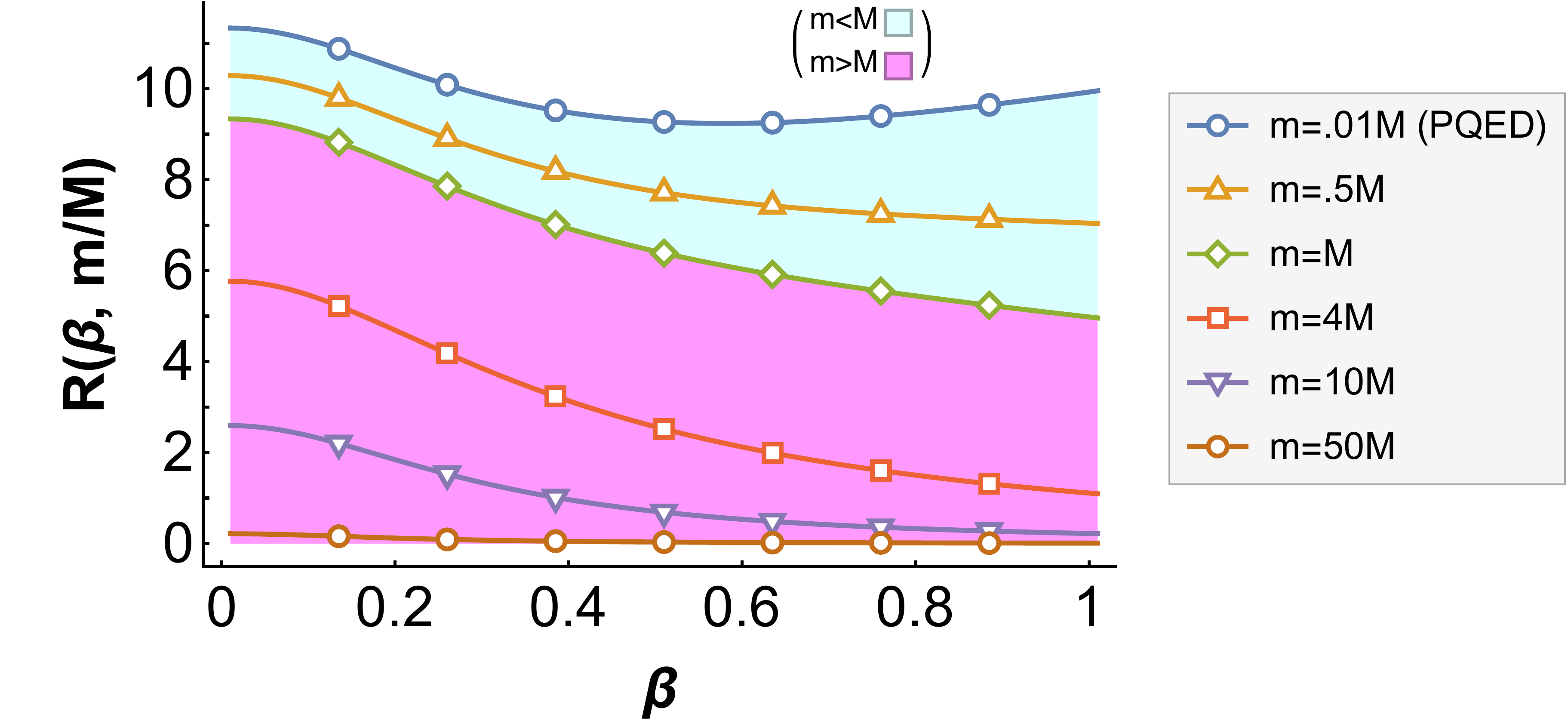}
    \caption{(Color Online). Dependence of $R$ on the ratio $m/M$ of the three-dimensional vector field mass to the electron mass. 
      {}For $(m/M)=0$, the PQED results are
      retrieved and for $(m/M)\gg 1$ the \textit{g}-factor is vanishingly small.
      The solid blue line (obtained near the  limit of non-massive gauge field) is related to PQED and gives the approximate values of $R$ to correct the  \textit{g}-factor  obtained in Ref.~\cite{fator-g}.
      }
    \label{CorrectionNatalia}
\end{figure}
 \begin{figure}[h]
    \centering 
   \includegraphics[width=7.6 cm]{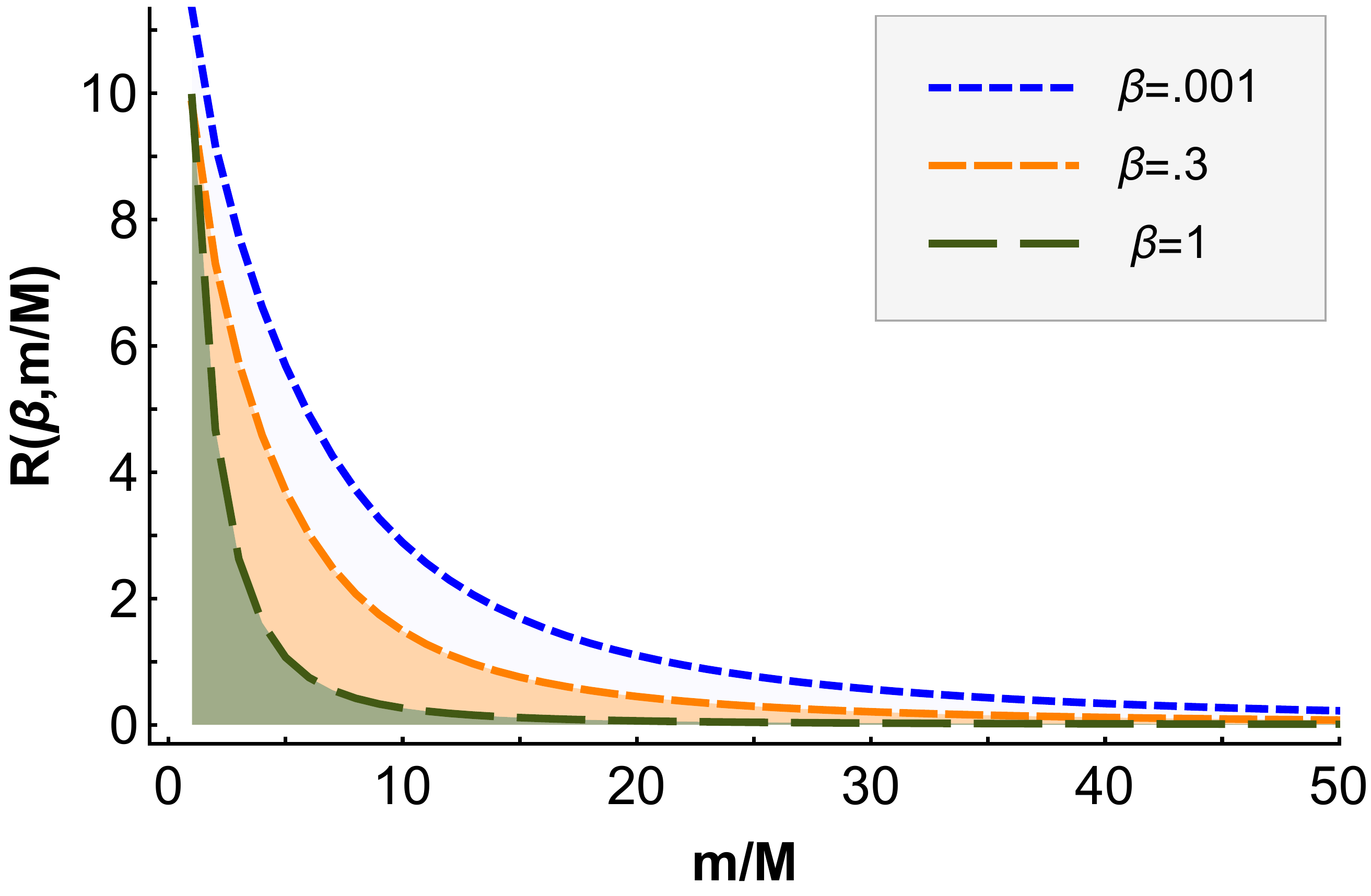}
    \caption{(Color Online). Dependence of ${R}$ on the ratio ($m/M$) for some values of $\beta$ obtained numerically. We notice that the vector field mass suppresses the result of the integral (and consequently of the $g$-factor) for all values of $\beta$, but does it faster for ultra-relativistic systems ($v_F\approx c$).
    }
    \label{MassDependence}
\end{figure}

{As a consistency check, we take the limit of Eq.~(\ref{Rbar}) when $(m/M) \to 0$ and obtain
\begin{gather} \displaystyle
R=
\int\displaylimits_{0}^{1} d x \int\displaylimits_{0}^{1-x}\frac{dy}{\sqrt{1-x-y}
}
\frac
{
2+4\beta^2+\beta^2(x+y)
}
{ (x+ y) [1-(1-\beta^2)\,(x+y) ]}\label{RbarNat},
\end{gather}
which has a similar structure to the parametric integral in Ref.~\cite{fator-g}, except that the term 2 in the numerator of Eq.~(\ref{RbarNat}) was missing  in that reference.
 Hence, our result not only extends the projection to the massive case, but also corrects the \textit{g}-factor result for graphene found previously in Ref.~\cite{fator-g}.} 

{
Although we could not find an analytical solution for the integral in Eq.~(\ref{Rbar}),  some numerical solutions for $R$ are shown in Fig.~\ref{CorrectionNatalia} for different mass ratios ($m/M$).
These  numerical values of $R$, when substituted into Eq.~(\ref{F2}), allow us to verify how small changes in $\beta$ affect the \textit{g}-factor in each case.

To show that the \textit{g}-factor associated with the system is going to decrease faster with the mass when $v_F$ gets closer to $c$, we plot
in Fig.~\ref{MassDependence} the numerical values of ${R}$ for a few different values of $\beta$, but now as a function of the mass ratio ($m/M$).
}

\section{Conclusions}
\label{Resultados}

In this work, we have provided a generalization of the dimensional
reduction, developed in Ref.~\cite{Marino1}, for the case of a
(3+1)D \textit{massive} vector field. By doing so, we have
obtained an effective planar model that retains the fundamental
physical properties of the Proca electrodynamics, which is taken as an 
effective model for describing massive (via the Anderson-Higgs-Meissner mechanism)
 photons in a material.
 We have then
evaluated the one-loop radiative corrections for the electron and
vector field self-energies for this model. 

We observed that the divergent
part of the electron self-energy is exactly identical to the one
obtained in the context of the PQED (massless) model. Therefore, from
the renormalization group perspective, the renormalization of the
{}Fermi velocity does not depend on the mass of the vector field and,
consequently, it is identical  to the one obtained for the PQED
model~\cite{Voz2,livroMarino}, even though the effective mass for the electron is seen
to increase with the vector field mass.

{
For the vertex diagram, we have found that the parametric integral in its form factor ---
and thus this theory's predicted \textit{g}-factor --- decreases with both the value of $\beta$ and the ratio ($m/M$), effectively switching  the magnetic coupling off of the system for very large values of $m$.

This work contributes to the proper description of planar effective models, which are known to be intrinsically related to condensed-matter physical systems, by investigating the effects of attributing a mass to the interaction field.
}
\acknowledgements

R.F.O. is partially supported by Coordena\c{c}\~ao de
Aperfei\c{c}oamento de Pessoal de N\'{\i}vel Superior -- Brasil
(CAPES), finance code 001, and by CAPES/NUFFIC, finance code 0112;
V.S.A. and L.O.N. are partially supported by research grants from Conselho
Nacional de Desenvolvimento Cient\'{\i}fico e Tecnol\'ogico (CNPq) and
by CAPES/NUFFIC, finance code 0112;  V.S.A. acknowledges the Institute
for Theoretical Physics of Utrecht University for the kind hospitality
and M. Gomes for very fruitful discussions; E.C.M. is partially
supported by both CNPq and Funda\c{c}\~ao Carlos Chagas Filho de
Amparo \`a Pesquisa do Estado do Rio de Janeiro (FAPERJ).  R.O.R is
partially supported by research grants from CNPq, grant
No. 302545/2017-4 and FAPERJ, grant No.  E --
26/202.892/2017. J.F.M.N. and R.F.O. are also grateful to M.C. Lima and G.C. Magalh\~{a}es for interesting discussions about the model.

\appendix
\section{Vertex correction}
\label{app}

To shed more light on the \textit{g}-factor corrections that come from
the vertex developed in
Subsec.~\ref{photon_mass_in_vertex_correction}, we make the steps
that differ from the usual QED more explicit. The subtlety on the
calculation that appears for the model discussed here comes from the
fact that it is anisotropic and has a different photon propagator (the
planar projection of a massive vector field). 

Starting from the vertex structure in
Subsec.~\ref{photon_mass_in_vertex_correction}, 
\begin{equation} \Gamma^{\nu}_{\rm 1L} =   \int
  \frac{d^D k}{(2\pi)^D} \Gamma^{\alpha} \underbrace {S_F(k +
    p')\Gamma^{\nu} S_F(k + p)}_{\rho^\nu}
  \Gamma^{\beta}\Delta_{\alpha\beta}(k) , 
\end{equation}
we define an auxiliary variable $\rho^\nu$ and expand
 the indices $\alpha$ and $\beta$ to get
\begin{equation}\label{rhonu}
\Gamma^{\nu}_{\rm 1L} =   \int \frac{d^D k}{(2\pi)^D} \left[
  \Gamma^{0} {\rho^\nu} \Gamma^{0}\Delta_{00}(k)+    \Gamma^{l}
        {\rho^\nu} \Gamma^{n}\Delta_{ln}(k)\right]. 
\end{equation}
In Eq.~(\ref{rhonu}), the non-diagonal terms of the vector field propagator
were dropped because the metrics inside them vanish. Explicitly
substituting the Feynman rules given in Sec.~\ref{FRules}, we find
\begin{widetext}
\begin{gather}\nonumber
\hspace{-14cm}\Gamma^{\nu}_{\rm 1L} = - \frac{ie^2 c}{2}\mu
\displaystyle\int \frac{d^D k}{(2\pi)^D} \\\left\{ \gamma^{0} \left(
\dfrac{\gamma^0 (k + p')_0 + v_F \gamma^i (k + p')_i + M c^2}{(k +
  p')_0^2-v_F^2 {\bf (k + p')}^2-M^2 c^4} \right)\right.
\Gamma^{\nu}  \left( \dfrac{\gamma^0 (k + p)_0 + v_F \gamma^i (k +
  p)_i + M c^2}{(k + p)_0^2-v_F^2 {\bf (k + p)}^2-M^2 c^4} \right)
\gamma^{0} \left[\frac{1} {\sqrt{k_0^2 - c^2 \, {\mathbf k}^2 - m^2
      c^4}}\right]
+\nonumber\\ \beta^2\gamma^{l} {\left( \dfrac{\gamma^0 (k +
    p^\prime)_0 + v_F \gamma^i (k + p')_i + M c^2}{(k +
    p^\prime)_0^2-v_F^2 {\bf (k + p')}^2-M^2 c^4} \right)
}{ \Gamma^{\nu}  \left( \dfrac {\gamma^0 (k + p)_0 + v_F \gamma^i (k +
    p)_i + M c^2} {(k + p)_0^2-v_F^2 {\bf (k + p)}^2-M^2 c^4} \right)
}\left.  \gamma^{n} \left[\frac{\eta_{ln}} {\sqrt{k_0^2 - c^2 \,
      {\mathbf k}^2 - m^2 c^4}}\right] \right\}.\label{Lengthy}
\end{gather}
\end{widetext}

To handle this unusual photon propagator, which contains a square root and a mass term, we apply a slightly modified Feynman
parameterization,
\begin{equation}\label{Fparam}
\frac{1}{a b d^{{1}/{2}}}=\frac{3}{4} \int_{0}^{1} d x \int_{0}^{1-x}
d y \frac{(1-x-y)^{-{1}/{2}}}{[a x+b y+d(1-x-y)]^{{5}/{2}}}
\end{equation}
on the denominator of each term in Eq.~(\ref{Lengthy}). Assuming \vspace{-.5cm}
\begin{eqnarray*}
a&=&{(k + p')_0^2-v_F^2 {\bf (k + p')}^2-M^2 c^4},\\ b&=&{(k +
  p)_0^2-v_F^2 {\bf (k + p)}^2-M^2 c^4},\\ d&=&3{{k_0^2 - c^2 \,
    {\mathbf k}^2 - m^2 c^4}},
\end{eqnarray*}
we find that the denominator of Eq.~(\ref{Lengthy}) is given by
\begin{gather*}\label{Fparamed}
\frac{1}{a b c^{1/2}}=\frac{3}{4} \int_{0}^{1} d x \int_{0}^{1-x} d y
\ (1-x-y)^{-1/2}\\ \{ k_0^2(1+x+y)+[ 2k_0 p_0^\prime + {p^\prime_0}^2
] x \\ [{2k_0 p_0+ p_0^2  - v_F^2 {\mathbf (k + p)}^2  - M^2 c^4}
]y+\\ - c^2 \, {\mathbf k}^2 - m^2 c^4 (1-x-y)\}^{-5/2}.
\end{gather*}

Now, we complete the $k_0$ square, make a shift $k_0\to k_0- \omega_0$
(with $\omega_0 = x\,p_0' + y\,p_0$) on it, and solve the integration
over $k_0$ as done in standard anisotropic procedures (at this point,
we also drop the odd terms in $k_0$). 

Then, we follow essentially the same steps for $\mathbf{k}$, but using
the shift $\mathbf{k}\to\mathbf{k}-\text{ \boldmath
  $\omega$}{\beta^2}/{\cal D}^2$, with ${\cal D}
=[(1-\beta^2)\,(x+y) -1]$ and  $\text{ \boldmath $\omega$} =
x\,\mathbf{p'} + y\,\mathbf{p}$. {}Finally, we perform the integration
over $\mathbf{k}$ using dimensional regularization \cite{Peskin} to
obtain Eq.~(\ref{vertice}), namely:
\begin{equation}
\Gamma^{i}_{\rm 1L} = \int_0^1{dx} \int_0^{1-x} \dfrac{dy}{6
  \pi^2}  \left[\dfrac{N_3}{\cal K}\right]  \dfrac{M\, p^i
  v_F^2}{\sqrt{1-x-y}},
\end{equation}
where, for convenience, we defined the numerator result for the
integration in $\mathbf{k}$  as 
\begin{gather*}
N_3 =\left.\frac{ 2(1-x-y) \gamma^j \omega_j p_i v_F^2 }{{\cal
    D}}\right.  -\left.\frac{[\omega_i-2(x+y)^2 p_i] \beta^4}{{\cal
    D}^2}\right.,	\nonumber \\ {\cal D} =[(1-\beta^2)\,(x+y)
  -1],
\end{gather*}
and its denominator as 
\begin{gather*}
{\cal K} ={\cal D}^2 \left(M^2 c^4-m^2c^4-\omega_0^2\right) c^2 +
\left[v_F^2 (p'-p)_i^2+p_0^2\right] x y		\nonumber \\
+\left[(x+y)(1-x-y)c^2\right] (p'+p)_\mu^2	 +p_0^2 y^2  .
\end{gather*}

Using the (Gordon) decomposition identity
\begin{equation}	\nonumber
 \Bar{u}(p')  \gamma ^\mu u(p) =\Bar{u} (p')
 \left[\frac{(p'+p)^\mu}{2Mc^2}+ i \sigma^ {\nu \mu} \frac{ q_\nu}
   {2Mc^2} \right]\,u (p)  ,
\end{equation}
where $\Bar{u}(p')$ and $u(p)$ are spinorial solutions to the Dirac
equations,  $q_\nu$ is the transferred momentum given by $q_\nu = (p'
- p)_\nu$, and    $\sigma_{\nu \mu}$ is defined as $\sigma_{\nu
  \mu}=(i/2)\left[\gamma_\mu,\gamma_\nu\right]$, we stablish which of
the terms of $M^i$ will be relevant to the \textit{g}-factor. In other
words, we select only the terms proportional to $\sigma^{i \alpha}$
from Eq.~(\ref{vertice}) and leave aside the terms that would have a
$(p'+p)^\mu$ to define 
\begin{equation}
\begin{split}\nonumber
M^{i}_g = \Bar{u}(p')\left\{ \int_0^1 dx \int_0^{1-x} \frac{dy}{6
  \pi^2} 	\frac{\;M\, v_F \, \sigma^{i\alpha} (p' -
  p)_\alpha\beta^2\textbf{}}{{\cal K} \sqrt{1-x-y}}\right.  \\\left.
\left[ {\dfrac{2 (x+y)}{\cal D}}-{\dfrac{\beta^2(x+y)^2}{{\cal
        D}^2}} \right]\right\} u(p),
\end{split}
\end{equation}
 as it appears in  Eq.~(\ref{integF2})
%


\end{document}